\documentclass[10pt,a4paper,tightenlines,twocolumn,prd,showkeys,lengthcheck]{revtex4}
\usepackage{amsfonts}
\usepackage{amsmath}
\usepackage{amssymb}
\usepackage{epsfig}
\usepackage{graphicx}
\usepackage{fancyhdr}

\begin{document}

\preprint{}
\title[ ]{\textbf{Towards a dynamical theory of observation}}
\author{George Jaroszkiewicz}
\affiliation{University of Nottingham, University Park, Nottingham, NG7 2RD, UK}

\begin{abstract}
We
introduce a model of classical and quantum observation based on contextuality and
dynamically evolving apparatus. Power sets of classical bits model the four
classical states of elementary detectors, viz. the two normal yes/no signal
states, the faulty or decommissioned state and the non-existence state. Operators over power set
registers are used to describe various physical scenarios such as the
construction and decommissioning of physical devices in otherwise empty laboratories, the
dynamics of signal states over those detectors, the extraction of
information from those states, and multiple observers. We apply our quantum formalism to the Elitzur-Vaidman 
bomb-tester experiment and the Hardy paradox experiment. 
\end{abstract}

\maketitle

\section{Introduction}

The Nineteenth and Twentieth Centuries may be summarized as the centuries of
non-relativistic classical science and relativistic quantum science
respectively. Judging by the scale of activity and progress in the field,
the Twenty-First Century may well turn out to be the century of neural
science. The hard problem of consciousness and its relationship to brain
function is being systematically worn down using all the quantum
technologies and theories developed over the last hundred years.

Central to this programme is the concept of the \emph{observer}, the
enigmatic `\emph{I}' of \emph{I think therefore I am}. The problem is that,
despite the many triumphs of quantum mechanics, the physics of the observer
and observation is still not well understood. An important problem is that
we are not sure what the correct way to model observers mathematically is.
The \emph{exophysical} perspective, which assumes that observers stand
outside the space-time arena in which SUOs (systems under observation)
exist, remains the dominant paradigm in all the Sciences.

Reality is different however. The empirical facts are that actual observers
are part of the physical universe and can be observed by other observers. As
Feynman wrote \cite{FEYNMAN-1982}: `\emph{... we have an illusion that we can do any
experiment that we want. We all, however, come from the same universe, have
evolved with it, and don't really have any real freedom. For we obey certain
laws and have come from a certain past.}'

This raises a question central to the interpretation of quantum mechanics:
are observers just more complicated versions of SUOs, a view of reality
known as the \emph{endophysical} perspective, or are they fundamentally
different altogether? This question has been called the endo-versus-exo
debate.

Another debate of central importance to the theory of observation involves
the conflict between the classical world view and the quantum world view.
The former postulates that SUOs and their properties exist independently of
any observers or observation, whereas the latter cannot make sense without
them.

Regardless of how observers are defined and whether classical or quantum
principles are involved, physicists generally believe that classical
information in some form is extracted from SUOs in actual physics
experiments. In all branches of science, their language reflects this
belief. Experimentalists talk of measuring an electron's spin or the mass of
a new particle, and so on. Whilst this point of view is of immense practical
value, it may be a fundamental conceptual error. To quote Heisenberg \cite{HEISENBERG-1927}:
`\emph{the orbit} [of the electron] \emph{comes into being only when we
observe it}.' It is difficult to think of any idea further from the
classical world view than that one.

The conceptual issues in quantum mechanics such as wave-particle duality,
quantum interference and non-locality gave the first indication that all
might not be well with this perspective. We need only to look at the photon
concept to appreciate some of the problems with the idea that photons are
particles \cite{PAUL-2004}.

Developments in neuroscience are reinforcing the need to rethink the
universality of the exophysical perspective. If large collections of neurons
are seen to act in a coherent fashion more typical of observers than SUOs,
then the boundaries between endo and exo physics will have been eroded. Just
when does a brain become an observer rather than an SUO?

What seems to be missing is a dynamical theory of observation which regards
observers and SUOs more on the same footing. In such a theory, observers
would be subject to the same laws of physics as the SUOs that they were
observing, just as Feynman said in 1982. Such a theory would ideally be
capable of accounting for the creation and annihilation of observers and
their apparatus, because in the real world, nothing lasts forever.

We are a long way from having such a theory of observation, but various authors have made 
some interesting comments on this topic \cite{ANASTOPOULOS-2006}. This paper
outlines our current thoughts on the subject to date. These are based firmly
on standard principles of quantum mechanics extended to cover the processes
of observation. The most important idea is to discuss observation in terms
of signal states of apparatus, referred to as \emph{labstates}, rather than
states of SUOs.

\section{Contextuality}

When we look more closely at the meaning of observation, some concepts begin
to emerge as more fundamental than others. Perhaps the most important of all
is \emph{contextuality}, the idea that every statement is true or physically
meaningful only relative to its context. Outside of that context, nothing
can or should be said about the truth or validity of that statement.

Contextuality makes sense of Heisenberg's remark about electron
trajectories, quoted above. We cannot say that electron trajectories exist
or otherwise if we do not observe them. We should say nothing in that
context. Contextuality also helps settle the particle-wave issue in quantum
mechanics: there is no context in which an SUO can appear to be completely a
particle and completely a wave.

A statement is \emph{absolute} if it is true regardless of context. A
central assumption of the classical world view is that there are mechanical
absolutes. Newton recognized the need to clarify this point and took care to
make specific statements about Absolute Space and Absolute Time in his
monumental book \emph{The Principia} \cite{NEWTON-1687}. The problem is that quantum
mechanics has no mechanical absolutes: wave-functions are contextual.

Contextuality leads to our first principle of quantum observation:

\ 

\noindent \textbf{Principle I:} \emph{There are no absolutes in physics}.

\ 

With a veto on the absolute, contextuality enforces a better mental attitude
towards important contextual concepts in physics such as entropy and
probability. Contextuality makes us wary of formalisms which treat quantum
wave-functions in absolute terms, as in Bohmian mechanics \cite{BOHM-1952}, the
Multiverse paradigm \cite{DEUTSCH-1997} and some versions of decoherence theory
\cite{ZUREK-2002}.

Contextuality gives an answer to the EPR debate \cite{EPR-1935}, which is
essentially about observation in quantum mechanics. We have to reject the
concept of `element of reality', slipped into the EPR paper as a seductive
appeal to reason. If we subscribe to that idea then we are being manipulated
into accepting the classical view that SUOs can have absolute properties.
Such thinking is inconsistent with Principle I stated above.

Contextuality and the concept of an \emph{equivalence class} leads to a
useful definition of what is meant by an SUO. Recall that an equivalence
class is a subset of a set, such that all elements of that subset have some
property or properties in common. Those properties provide the context for
the definition of that equivalence class. Spekkens \cite{SPEKKENS-2005} 
gives a clear discussion of these concepts as they relate to measurements. 

In our approach, SUOs are discussed in terms of equivalence classes defined
by the physics and context of a given observation, an approach used by
Ludwig \cite{LUDWIG-1983} and Kraus \cite{KRAUS-1983} in their work on the quantum measurement
problem. We define \emph{relative external context} to be those properties
of and that information about the state of an SUO which is washed out, or
redundant, in the definition of an equivalence class, whilst \emph{relative
internal context} refers to the defining properties of an equivalence class
associated with a given experiment. For example, in the measurement of
electron spin in the Stern-Gerlach experiment \cite{SG-1922}, the momentum and
position of the electron plus whatever is happening in the rest of the
universe represent the relative external context whereas whether the
electron is in its spin up or down state is internal context, relative to
that particular experiment.

The equivalence class approach leads to a second principle of quantum
observation, implicit in all experimental science:

\ 

\noindent \textbf{Principle II:} \emph{Relative external context can be
ignored in any experiment}.

\ 

In standard quantum mechanics, this principle applies to state preparation
as well as outcome detection. Peres made the point that a prepared quantum
state of a system carries no memory of how it was prepared \cite{PERES-1993}.

A point in favour of the equivalence class perspective is that it accords
with the spirit of Heisenberg's remark about electron trajectories quoted
above, since it is clearly the apparatus which defines the equivalence
classes being observed.

\section{Bits and qubits}

\emph{Classical bits} are central to our approach, being used to represent
the process of observation in its simplest possible form. A classical bit $%
\mathsf{B}\equiv \{0,1\}$ is a set with two distinct elements, called \emph{%
classical bit states}, plus a context which gives those states a meaning.
Bit states are used to label equivalence classes of SUOs for those
situations where only two alternatives exist as far as the observer is
concerned at that time.

In recent years, the quantum analogue of a bit, known as a qubit, has found
many uses, particularly in quantum computation \cite{NIELSEN+CHUANG-2000}. A
qubit is altogether a more complicated mathematical object than a bit. One
important difference is that whereas bits are not vector spaces, qubits are
complex vector spaces, which means that elements of a qubit can be
multiplied by complex numbers and can be added together. Another important
difference is that a qubit contains a zero vector, which a bit does not. We
will use the power set of a bit to get around that particular point.

It will be useful to us later to briefly review some basics aspects of a
qubit now. Elements of a qubit $Q$ and its dual $Q^{\ast }$ are denoted by
ket $|\psi \rangle $ and bra $\langle \psi |$ vectors respectively. Given
orthonormal bases $\{|0\rangle ,|1\rangle \}$ and $\{\langle 0|,\langle 1|\}$
for a qubit and its dual respectively, we define the \emph{projection
operators} $p^{0}\equiv |0\rangle \langle 0|$, $p^{1}\equiv |1\rangle
\langle 1|$ and the \emph{transition operators}\textbf{\ }$a\equiv |0\rangle
\langle 1|$, $a^{+}\equiv |1\rangle \langle 0|$. These operators satisfy the
product rules given in Table $1$.

\begin{table}
\caption{The product table for the four basic qubit operators, where $0$
represents the zero operator. Entries in the main square represent products
of left-most column elements with top row elements in that order.}
\label{tab:1}
\begin{center}
\begin{tabular}{c|cccc}
& $p^{0}$ & $p^{1}$ & $a$ & $a^{+}$ \\ \hline
$p^{0}$ & $p^{0}$ & $0$ & $a$ & \multicolumn{1}{c|}{$0$} \\ 
$p^{1}$ & $0$ & $p^{1}$ & $0$ & \multicolumn{1}{c|}{$a^{+}$} \\ 
$a$ & $0$ & $a$ & $0$ & \multicolumn{1}{c|}{$p^{0}$} \\ 
$a^{+}$ & $a^{+}$ & $0$ & $p^{1}$ & \multicolumn{1}{c|}{$0$} \\ \cline{2-5}
\end{tabular}%
\end{center}
\end{table}

\section{Elementary signal detectors}

Our strategy is not to think of the SUOs as if they were `there' but to
discuss only that information which an observer can extract from \emph{%
elementary signal detectors}, or ESDs. Each ESD has only two possible normal
states, known as the \emph{ground} \emph{state} and the \emph{signal} \emph{%
state} respectively. Whenever an observer looks at a normally functioning
ESD, they will find it only in one of these two possible states, denoted by $%
0$ for the ground state and $1$ for the signal state.

Our approach assumes that any observation can be described in terms of
collections of ESDs. How many ESDs are needed in any particular experiment
will depend on context. Some experiments, particularly many in quantum
optics, can be described with a relatively small number of ESDs, whilst
other experiments may require enormous numbers. An example of the latter
type of experiment is the double-slit experiment, where we would require
very many ESDs to model all the positions on the detecting screen where a
photon could be detected.

ESDs are not restricted to the detection of position in space at a given
time. An ESD is any process of observation which will return either a \emph{%
yes} or a \emph{no} answer. In principle this could involve a great deal of
spatially extended physical equipment operated over relatively long periods
of time. Recent experiments in quantum optics such as the quantum eraser
\cite{WALBORN-2002} and delayed choice experiments \cite{JACQUES-2007} 
have reinforced the message that the process of observation in
quantum mechanics can appear not to follow classical patterns of causality
or locality \cite{KIM-2003}.

A typical experiment will involve a time-dependent collection of ESDs. It
would normally be assumed that prior to any run (or repetition) of the
experiment, each ESD would have been set in its characteristic ground state.
This has nothing to do with energy. The ground state of an ESD is simply
whatever condition the observer regards the ESD as having in the absence of
a response to any external stimulus.

If subsequently during the act of observation an ESD were found still in its
ground state, that would be taken as indicating that nothing had happened at
that ESD. If on the other hand an ESD were found in its signal state, then
something must have happened there, such as a particle impacting on a
detector.

There are two important caveats to this interpretation which play a crucial
role in the formalism: i) an ESD might not exist, or ii) it might exist but
be faulty or decommissioned. These will be discussed in detail below.

\section{Bit power sets}

We will identify the two possible normal signal states of a functioning ESD
as the two elements of a bit. As we have mentioned, however, bits are not
vector spaces and there seems to be no meaning to the addition of bit state $%
0$ to bit state $1$, or even of the multiplication of a bit state by a real
or complex number.

There is in fact a way of defining bit state addition, of a kind, in terms
of set theory. We recall that the \emph{power set} $\mathcal{P}(S)$ of a set
is the set of all possible subsets of $S$ including the empty set $\emptyset 
$ and $S$ itself. The power set $\mathcal{P}(\mathsf{B})$ of a bit $\mathsf{B%
}$ therefore has four distinct elements: $\mathcal{P}(\mathsf{B})=\{|0),|1),|%
\mathsf{B}),|\emptyset )\}$, where we define $|0)\equiv \{0\}$, $|1)\equiv
\{1\}$, $|\mathsf{B})\equiv \{0,1\}$ and $|\emptyset )\equiv \{\emptyset \}$%
. In this scheme, $|\emptyset )$ is a non-trivial element of $\mathcal{P}(%
\mathsf{B})$ and counts as one element of the power set.

We shall work in terms of the elements of $\mathcal{P}(\mathsf{B})$ rather
than with the elements of $\mathsf{B}$ itself, identifying elements $|0)$
and $|1)$ of $\mathcal{P}(\mathsf{B})$ as synonymous with bit states $0$ and 
$1$ of $\mathsf{B}$. The value of using $\mathcal{P}(\mathsf{B})$ rather
than $\mathsf{B}$ itself is that the elements of the former are sets, so we
can use the set properties of \emph{union} and \emph{intersection }to make
some interesting constructions analogous to those found in qubit theory.

\section{Interpretation}

Before we proceed further however, we need to resolve the following problem:
the power set $\mathcal{P}(\mathsf{B})$ of a bit $\mathsf{B}$ appears to
have too many elements. Logic suggests that only the elements $|0)$ and $|1)$
of the power set are actually needed. What can the elements $|\mathsf{B)}$
and $|\emptyset )$ represent?

We turn to the physics of observation to answer this question. An
observation of an ESD can be regarded as the acquisition of an answer to a 
\emph{binary question}, such a question being one with a \emph{yes} or \emph{%
no }answer. For example, we could ask the question $Q_{1}\equiv $ \emph{is
this ESD in its signal state? }If we looked and found it was in that state,
the answer would be \emph{yes} and so the state of the ESD would be
represented by the element $1$ of the corresponding bit. Conversely, if the
ESD was not found in its signal state, we would normally assume it was in
its ground state and therefore we would represent that situation by the
element $0$.

The matter is not as straightforward as it seems, however. A subtle issue
arises concerning two-valued logic as it applies to physics. Given an ESD,
there are two related binary questions. One is $Q_{1}$ and the other is $%
Q_{2}\equiv $ \emph{is this ESD\ in its ground state? } Logically, we would
assume $Q_{1}$ and $Q_{2}$ were conjugate questions, but physically this
need not be true. In an experiment, we could not always be certain that an
answer \emph{no} to $Q_{1}$ implies an answer \emph{yes} to $Q_{2}$. The
reason is that in any real experiment, $Q_{1}$ and $Q_{2}$ would be
questions asked at different physical locations, as in the Stern-Gerlach
experiment. We must be careful not to rely on unwarranted counterfactuality
when dealing with quantum physics. We should adhere as much as possible to
the following principle advocated by Wheeler \cite{WHEELER-1978} and 
Peres \cite{PERES-1993}:

\ 

\noindent \textbf{Principle III:} \emph{An experiment not actually done does
not count}.

\ 

This principle needs to be used carefully. We \emph{are} allowed, in fact 
\emph{required}, to superpose quantum amplitudes from signal detector
sources whenever we do \emph{not} know which source is the real one, i.e.,
when we have no which-way information. This leads to the fourth principle of
observation:

\ 

\noindent \textbf{Principle IV:} \emph{Quantum superposition occurs in the
absence of classical which-way information}.

\ 

The most well-known and potent demonstration of this principle is found in
Feynman's path integral formulation of standard quantum mechanics \cite{FEYNMAN+HIBBS-1965}. 
The same principle applies also to our formalism whenever the
observer chooses not to observe the signal status of particular ESDs
\cite{J2008R}.

A fundamental point concerns the existence of the ESDs themselves. Consider
what happens in a real laboratory in the execution of a given run of an
experiment. Before any observation of an ESD could be made for that run, the
observer would have had to make a decision to perform a reading on it.
Suppose the observer did make such a decision but was unaware that the ESD
never actually existed. In such a case, even if the observer had decided to make an observation, 
no possible answer $0$ or $1 $ could be found. This scenario
will be interpreted as corresponding to the empty set element $|\emptyset )$
of the power set $\mathcal{P}(\mathsf{B})$. In words, $|\emptyset )$
represents the answer \emph{yes} to the binary question \emph{is it true
that this ESD does not exist?} We shall call the element $|\emptyset )$ the 
\emph{empty state}.

With this possibility and the two `normal' possibilities of an ESD being in
its ground state or its signal state, we have accounted for three of the
four elements of the power set $\mathcal{P}(\mathsf{B})$. We account for the
fourth element $|\mathsf{B})$ as follows. Suppose that the ESD did exist and
was accessible to the observer but had a technical problem and gave
unreliable readings. Not all physical equipment works perfectly all the
time. The element $|\mathsf{B})\equiv |0)\cup |1)$ $=\{0,1\}$ will be taken
to represent such a scenario. Essentially, any answer that the observer
obtained when the ESD was in state $|\mathsf{B})$ would be known to the
observer to be uncertain or ambiguous and therefore unreliable. We shall
call the element $|\mathsf{B})$ the \emph{faulty state}. Another interpretation of the faulty state is that the ESD may have been \emph{decommissioned} for one reason or another and is no longer functioning as a normal ESD capable of transmitting information to other ESDs.

\section{Union and intersection}

Now that we have an interpretation of all four elements of the power set $%
\mathcal{P}(\mathsf{B})$ we can explore the consequences of this line of
thinking.

The elements of the power set $\mathcal{P}(\mathsf{B})$ are sets themselves
and therefore union and intersection are defined for them. These generate
the rules of a Boolean algebra, with $|\emptyset )$ playing the role of the
Boolean element $O$ and $|\mathsf{B)}$ playing the role of the Boolean
element $I$. In this context, union $\cup $ and intersection $\cap $ play
the roles of the idempotent, associative and commutative operations normally
denoted by $\vee $ and $\wedge $ respectively in the theory of Boolean
algebras. Every element $t$ of a Boolean algebra has a complement $\bar{t}$%
, such that $t\vee \bar{t}=I $ and $t\wedge \bar{t}=O$. In our case, $|\bar{%
\emptyset})=|\mathsf{B)}$, $|\mathsf{\bar{B}})=|\emptyset )$, $|\bar{0})=|1)$
and $|\bar{1})=|0)$.

\section{Questions and answers}

The idea that observation is the process of getting answers to certain
questions gives an insight into the essential difference between SUOs and
observers. SUOs do not ask questions whereas observers do.

We now apply this idea to bit power sets. We introduced the elements of a
bit as representing the answers to a binary question. Given $|0)\equiv \{0\}$
represents \emph{no} and $|1)\equiv \{1\}$ represents \emph{yes}, we denote
the particular binary question involved by $(1|$ and write $(1|0) = 0$, $%
(1|1) = 1$. In words, $(1|$ is the compound question: \emph{does this ESD
exist, and if so, is it working normally, and if so, is it in its signal
state?} The bit state $|1)$ returns a simultaneous \emph{yes} to all three
component sub-questions.

We noted above that the Boolean algebra of the power set $\mathcal{P}(B)$
consists of elements each of which has a complement. Likewise, binary
questions have their complements. The complement of the question $(1|$ will
be denoted by $(0|$, which is the binary question \emph{does this ESD exist,
and if so, is it working normally, and if so, is it in its ground state?}
Then we have the relations $(0|0) = 1$, $(0|1) = 0$.

By analogy we may introduce the questions $(\mathsf{B}|$ and $(\emptyset |$,
which have corresponding properties. We interpret $(\mathsf{B|}$ as the
question \emph{does this detector exist and if so, is it faulty?} In the
case of the empty state, it is more convenient to ask about non-existence
rather than existence, so we define $(\emptyset |$ as the question \emph{is
it true that the detector does not exist?}

The situation has now become more complicated than expected, because now we
have four binary questions asked of four power set states. An extension of
notation is called for. We define $|2)\equiv |\mathsf{B)}$ and $|3)\equiv |\emptyset )$, whilst 
$(2|\ \equiv (\mathsf{B|}$
and $(3|\ \equiv (\emptyset |$. Then all sixteen question and answer
relations are given by the rule%
\begin{equation}
(i|j)=\delta _{ij},  \label{000}
\end{equation}%
where $\delta _{ij}$ is the Kronecker delta.

We should comment further on relations (\ref{000}). We emphasize that a
non-existent ESD cannot actually give a physical signal. The statement $%
(0|3)=0$ should be interpreted as $``$\emph{if an ESD does not exist
at a place then it is not true that, if we looked, we would find a detector
in its ground state at that place}$"$. Similarly, the statement $%
(3|3)=1$ is equivalent to $``$\emph{if an ESD does not exist at a
place then it is true that, if we looked, we would not find an ESD at that
place}$"$.

At this stage the four questions $(i|$, $i = 0,1,2,3 $, look like the basis elements of a
dual vector space $V^{\ast }$ whilst the four answers $|j)$, $j = 0,1,2,3$, look like
the basis elements of a vector space $V$. For this reason we shall call the
elements $(i|$ the \emph{duals} of the $|i)$. However, there are
significant differences which we cannot expound on here, except to say that
questions do not have the same status as answers: there is
usually a temporal ordering relative to the observer, with questions being asked before answers can
be obtained.

Another difference between bit questions and answers is that whilst $(0|$
and $(1|$ can be thought of as mutual complements on account of their
physical interpretation, the same is not obviously the case as far as the
physics of $(2|$ and $(3|$ are concerned, apart from the fact that
an answer $``$yes$"$ to either tells us that the ESD can be ignored for information extraction purposes.

\section{Bit operators}

A \emph{bit operator} is any mapping from the power set $\mathcal{P}(\mathsf{%
B})$ back into the power set. Given an element $|s)$ of $\mathcal{P}(\mathsf{%
B})$ and a bit operator $O$ then we denote the value of the operator's
action on $|s)$ by $O|s)$. There is a total of $4^{4}=256$ different bit
operators and only a few will be of use to us.

A useful way of representing bit operators is via matrices. 
The elements $|i)$ of the power set $\mathcal{P}(\mathsf{B})$ may be
represented by column matrices $[i]$ given by
\begin{equation}
|0)  \rightleftharpoons \lbrack 0]\equiv \begin{bmatrix} 1 \\ 0 \\ 0 \\ 0\end{bmatrix},\ \ \
|1) \rightleftharpoons [1]\equiv \begin{bmatrix} 0 \\ 1 \\ 0 \\ 0\end{bmatrix},\ \ \ etc. 
\end{equation}
We represent
the action of bit operator $O$ on $|i)$ by the action of a \emph{bit
matrix} $[O]$ on a column matrix $[i]$, such that%
\begin{equation}
O|i)\equiv |Oi)\rightleftharpoons \lbrack O][i] \equiv [Oi].
\label{operator}
\end{equation}%
In this matrix representation the dual elements $(i|$ are
represented by the row matrices $(0| \rightleftharpoons %
\begin{bmatrix}
1 & 0 & 0 & 0%
\end{bmatrix}$, $(1| \rightleftharpoons  
\begin{bmatrix}
0 & 1 & 0 & 0%
\end{bmatrix}$, etc. This is consistent with the question and answer
relations (\ref{000}).

We can use the bit matrix representation to define the operational meaning
of the dyadics $|i)(j|$. These can then serve as formal basis
elements in the expansion of bit operators. Given a bit operator defined by (%
\ref{operator}), we can write it as the formal (dyadic) expression $O =
\sum_{i=0}^{3}|Oi)(i|$. Products of bit operators are defined in the
natural way: given bit operators $O_{1}$, $O_{2}$, we define their `product' 
$O_{2}O_{1}$ by its action on any element $|i)$ of the power set $%
\mathcal{P}(\mathsf{B})$ according to the rule $O_{2}O_{1}|i)\equiv
O_{2}\{O_{1}|i)\}$. This product rule is associative but not
commutative. To see this we note that products of two bit matrices are also
bit matrices, with the operator $O_{2}O_{1}$ being represented in the matrix
representation by the matrix product rule $[O_{2}O_{1}]=[O_{2}][O_{1}]$. The
result follows because matrix multiplication is associative but not
commutative.

The following bit operators turn out to be useful:

\ 

\noindent i) The \emph{identity} $I$ maps every element back into itself,
i.e., $I|s_{i})=|s_{i})$. Its matrix elements are given by the Kronecker
delta, viz., $[I]_{ij}=\delta _{ij}$.

\ 

\noindent ii) The \emph{annihilator }$Z$ maps any element $|i)$ of the
power set $\mathcal{P}(\mathsf{B})$ into the empty state $|\emptyset )$,
viz., $Z|i)=|\emptyset )$, $i = 0,1,2,3$, so its matrix representation is%
\begin{equation}
\lbrack Z]=\begin{bmatrix} 0 & 0 & 0 & 0 \\ 0 & 0 & 0 & 0 \\ 0 & 0 & 0 & 0
\\ 1 & 1 & 1 & 1\end{bmatrix}.
\end{equation}
The annihilator has a fundamental role in our theory: it represents the process of destroying and removing all traces of an already existing ESD.

\ 

\noindent iii) The \emph{relative projection operators} $P^{0}$ and $P^{1}$ have the
action%
\begin{eqnarray}
\begin{array}{cccc}
P^{0}|0)=|0), & P^{0}|\mathsf{B})=|\emptyset ), & \ P^{1}|0)=|\emptyset ), & 
P^{1}|\mathsf{B})=|\emptyset ), \\ 
P^{0}|1)=|\emptyset ), & P^{0}|\emptyset )=|\emptyset ), & \ P^{1}|1)=|1), & 
P^{1}|\emptyset )=|\emptyset ),%
\end{array}%
\end{eqnarray}%
so their matrix representations are%
\begin{eqnarray}
\lbrack P^{0}]=\begin{bmatrix} 1 & 0 & 0 & 0 \\ 0 & 0 & 0 & 0 \\ 0 & 0 & 0 &
0 \\ 0 & 1 & 1 & 1\end{bmatrix},\ \ \ [P^{1}]=\begin{bmatrix} 0 & 0 & 0 & 0
\\ 0 & 1 & 0 & 0 \\ 0 & 0 & 0 & 0 \\ 1 & 0 & 1 & 1\end{bmatrix}.
\end{eqnarray}%
These operators are idempotent, viz., $P^{0}P^{0}=P^{0},\ \ \
P^{1}P^{1}=P^{1}$ and \emph{orthogonal}, viz. $P^{0}P^{1}=P^{1}P^{0}=Z$. In
this context, the annihilator $Z$ plays the role of a zero element.

\noindent iv) The \emph{signal creation }and\emph{\ signal annihilation }%
operators $\bar{A}$, $A$ are defined principally by their action on the
normal states $|0)$ and $|1)$: 
\begin{eqnarray}
\begin{array}{cccc}
A|0)=|\emptyset ), & A|\mathsf{B})=|\emptyset ), & \ \bar{A}|0)=|1), & \ \ 
\bar{A}|\mathsf{B})=|\emptyset ), \\ 
A|1)=|0), & A|\emptyset )=|\emptyset ), & \ \bar{A}|1)=|\mathsf{\emptyset )},
& \ \bar{A}|\emptyset )=|\emptyset ),%
\end{array}%
\end{eqnarray}
which gives the matrix representations%
\begin{eqnarray}
\lbrack A]=\begin{bmatrix} 0 & 1 & 0 & 0 \\ 0 & 0 & 0 & 0 \\ 0 & 0 & 0 & 0
\\ 1 & 0 & 1 & 1\end{bmatrix},\ \ \ [\bar{A}]=\begin{bmatrix} 0 & 0 & 0 & 0
\\ 1 & 0 & 0 & 0 \\ 0 & 0 & 0 & 0 \\ 0 & 1 & 1 & 1\end{bmatrix}.
\end{eqnarray}%
These operators are \emph{nilpotent}, viz. $AA=\bar{A}\bar{A}=Z$, with $Z$
once again playing the role of a zero element.

The product rules for the operators $P^{0}$, $P^{1}$, $A$, $\bar{A}$ are
given in Table $2$. Comparison with Table $1$ shows that these tables are
isomorphic, provided the zero operator in Table $1$ is identified with the
annihilator $Z$ in Table $2$. 
\begin{table}
\caption{The product table for the four basic bit operators, where $Z$
represents the annihilator. Entries in the main square represent products of
left-most column elements with top row elements in that order.}
\label{tab:2}
\begin{center}
\begin{tabular}{c|cccc}
& $P^{0}$ & $P^{1}$ & $A$ & $\bar{A}$ \\ \hline
$P^{0}$ & $P^{0}$ & $Z$ & $A$ & \multicolumn{1}{c|}{$Z$} \\ 
$P^{1}$ & $Z$ & $P^{1}$ & $Z$ & \multicolumn{1}{c|}{$\bar{A}$} \\ 
$A$ & $Z$ & $A$ & $Z$ & \multicolumn{1}{c|}{$P^{0}$} \\ 
$\bar{A}$ & $\bar{A}$ & $Z$ & $P^{1}$ & \multicolumn{1}{c|}{$Z$} \\ 
\cline{2-5}
\end{tabular}%
\end{center}
\end{table}

\noindent v) The \emph{construction operator }$C$ acts on every element $%
|i)$ of the power set and sets it to the ground state in readiness for
observation, i.e., $C|i)=|0)$, $i = 0,1,2,3$. There are two scenarios. If the bit is in
its empty state then its ESD does not exist, so the action of the
construction operator represents the physical construction of a standard ESD
in its ground state in the laboratory, prior to any experiment. It is
assumed that facilities exist in the laboratory for this. Alternatively, if
the ESD already exists, then the construction operator resets it to its
ground state if it is normal or repairs it and sets it to its ground state
if it is faulty. This operator is represented by the matrix%
\begin{equation}
\lbrack C]=\begin{bmatrix} 1 & 1 & 1 & 1 \\ 0 & 0 & 0 & 0 \\ 0 & 0 & 0 & 0
\\ 0 & 0 & 0 & 0\end{bmatrix}.
\end{equation}

\ 

\noindent vi) The \emph{decommissioning operator }$D$ represents the action
of decommissioning an already existing ESD, setting it into its faulty state $%
|2)$. This operator does \emph{not} reset states $|0), |1)$ and $|2)$ to the
non-existence state $|3)$ because in the real world, there will invariably
be some remaining information in the form of \emph{debris} which will inform
the observer that apparatus has been decommissioned. This is an important feature
 of our discussion towards the end of this paper of
the Elitzur-Vaidman bomb-tester experiment and Hardy's paradox experiment.

The decommissioning operator is represented by the matrix%
\begin{equation}
\lbrack D]=\begin{bmatrix} 0 & 0 & 0 & 0 \\ 0 & 0 & 0 & 0 \\ 1 & 1 & 1 & 0
\\ 0 & 0 & 0 & 1\end{bmatrix}.
\end{equation}

\section{Discrete time and space}

Classical mechanics generally assumes that SUOs move around continuous space
continuously, but we take a different view. How observers interact with SUOs
is always described in discrete terms, because real experiments always
involve data extraction occurring at discrete times on finite numbers of
ESDs. There are no truly continuous observations, either spatially or
temporally. Some experiments do deal with continuous variables, such as
temperature, but this is conceptually very different to what is being
discussed here and will not be considered further in this article.

We do not refer to Hamiltonians or continuous unitary evolution either.
Quantum dynamical evolution is discussed in terms of mappings from one
quantum register to another and the unitary evolution operators of standard
quantum mechanics are replaced by semi-unitary operators \cite{J2008R}.

A typical experiment described by our formalism involves an observer
interacting with a time-dependent number $r_{n}$ of ESDs at a countable
number of times $t_{n}$, where the integer $n$ runs from some initial
integer $M$ to some final integer $N>M$. There is no need to assume that $%
t_{n+1}-t_{n}$ always has the same value, or that $r_{n}$ is independent of $%
n$. The formalism allows for the creation and destruction of ESDs, something
which happens in the real world.

\section{The laboratory and the universal register}

In our approach it is assumed that an observer exists in a physical
environment referred to as the \emph{laboratory, }$\Lambda $. This will have
the facilities for the construction or introduction of apparatus consisting
of a number of ESDs. At any given discrete time $n$, the observer will
associate a state known as the \emph{labstate} to the collection of ESDs at
that time. This state could be a pure state or a mixed state. We shall
restrict our attention in this paper to pure labstates for reasons of space.

A labstate carries information as to whether various ESDs exist in the first
place and, if so, whether they are functioning normally and either in their
ground or signal states, or whether they are faulty.

The power set approach to ESDs allows us to think of an absence of an ESD in 
$\Lambda $ as an observable fact which can be representable mathematically.
The state corresponding to an absent or non-functioning ESD is represented
by the element $|\emptyset )$ of its associated power set. Therefore, we can
represent a complete absence of any ESDs whatsoever by an infinite
collection of such elements. This corresponds to an observer without any
apparatus, i.e., an empty laboratory. We denote this labstate by the symbol $%
|\Omega )$ and call it the \emph{information void}, or just the \emph{void}.
It represents a potential for existence, relative to a given observer.

If the observer's laboratory $\Lambda $ is in its void state $|\Omega )$,
that does not mean that the laboratory $\Lambda $ or the observer do not
exist, or that there are no SUOs in $\Lambda $. It means simply that the
observer has no current means of acquiring any information. An empty
laboratory devoid of any detectors is a physically meaningful concept, but
one with no interesting empirical content.

The information void can be thought of as one element in an infinite set
called the \emph{universal register }$\mho $, the Cartesian product of an
infinite number of bit power sets. We write%
\begin{equation}
|\Omega )\equiv \prod_{i}^{\infty }|\emptyset _{i})\in \mho \equiv
\prod_{i}^{\infty }\mathcal{P}(\mathsf{B}^{i}),  \label{xxx}
\end{equation}%
where the index $i$ could in principle be discrete, continuous or a
combination of both. The cardinality of the universal register as a measure
of how many power sets belong to it may be assumed to be infinity, but
precisely what sort of Cantorian cardinality it should be is not clear. If
we thought in terms of $\Lambda $ sitting in continuous space, then we
expect at least the cardinality, $\mathfrak{c}$, of the continuum. However,
that is a metaphysical statement, because there would not be enough energy
in the universe to create a continuum of ESDs \footnote{%
In other words, the universe cannot observe itself completely.}. Principle
II comes to our aid here. Real observers can only ever deal with finite
numbers of ESDs in practice and by Principle II we can generally ignore all
potential ESDs \footnote{%
An absence of an ESD could be significant in some circumstances however.}

The product notation in (\ref{xxx}) is not essential but has been chosen to
reflect the relationship between collections of power sets and the tensor
products of qubit spaces that we encounter in the quantum version of this
approach, discussed later. In our products, ordering is not significant,
since labels keep track of the various terms. An arbitrary classical labstate $|\Psi )$
in the universal register $\mho $ will be of the form $\prod_{i}^{\infty
}|s_{i})$, where $|s_{i})$ is one of the four elements of $\mathcal{P%
}(\mathsf{B}^{i})$.

Operators acting on universal register states will be denoted in blackboard
bold font and act as follows. If $O_{i}$ is a bit operator acting on
elements of $\mathcal{P}(B^{i})$, then $\mathbb{O\equiv }$ $%
\prod_{i}^{\infty }O_{i}$ acts on an arbitrary classical state $|\Psi )\equiv
\prod_{i}^{\infty }|s_{i})$ according to the rule%
\begin{equation}
\mathbb{O}|\Psi )\equiv \prod_{i}^{\infty }O_{i}|s_{i}).
\end{equation}

For every classical register state $|\Psi )\equiv \prod_{i}^{\infty }|s_{i})$
there will be a corresponding dual register state $(\Psi |\equiv
\prod_{i}^{\infty }(s_{i}|$, where $(s_{i}|$ is dual to $|s_{i})$%
. Classical register states including the void satisfy the orthonormality condition%
\begin{equation}
(\Phi |\Psi )\equiv \left\{ \prod_{i}^{\infty }(r_{i}| \right\}
\prod_{j}^{\infty }|s_{j}) =
\prod_{i}^{\infty }(r_{i}|s_{i})=\prod_{i}^{\infty }\delta _{r_{i}s_{i}}.
\end{equation}
Classical register states $|\Phi )$, $|\Psi )$ which differ in at least one bit power
set element therefore satisfy the rule $(\Phi |\Psi )=(\Psi |\Phi )=0$.

\section{Contextual vacua}

In conventional classical mechanics or Schr\"{o}dinger-Dirac quantum
mechanics, empty space is generally not represented by any specific
mathematical object. In quantum field theory, however, empty space is
represented by the \emph{vacuum}, a normalized vector in an infinite
dimensional Hilbert space. It has physical properties such as zero total
momentum, zero total electric charge, etc., which although bland are
physically significant attributes nevertheless.

In our approach we encounter an analogous concept. Starting with the void $%
|\Omega )$, we represent the construction of a collection of ESDs in the
laboratory $\Lambda $ by the application of a corresponding number of
construction operators $C_{i}$ to their respective empty states $|\emptyset
_{i})$. For example, a labstate consisting of a single ESD $i$ in its ground
state is given by $|\Psi )=\mathbb{C}_{i}|\Omega )=\left\{ \prod_{j\neq
i}^{\infty }|\emptyset _{j})\right\} \times |0_{i})$ where $\mathbb{C}_{i}$
is the register operator $\mathbb{C}_{i}\equiv \left\{ \prod_{j\neq
i}^{\infty }I_{j}\right\} \times C_{i}$. More generally, a state consisting
of a number $r$ of ESDs each in its ground state is given by%
\begin{equation}
|\Psi ^{r})=\mathbb{C}_{1}\mathbb{C}_{2}\ldots \mathbb{C}_{r}|\Omega ),
\label{lab}
\end{equation}%
where without loss of generality we label the ESDs involved from $1$ to $r$.
Such a state will be said to be a \emph{rank}-$r$ \emph{ground state}, or \emph{contextual vacuum state}.

We can now draw an analogy between the vacuum of quantum field theory and
the rank-$r$ ground states in our formalism. The physical three-dimensional
space of conventional physics would correspond to a ground state of
extremely large rank, \emph{if} physical space were relevant to the experiment.
This would be the case for discussions involving particle scattering or
gravitation, for example. For many experiments however, such as the
Stern-Gerlach experiment and quantum optics networks, physical space would
be considered part of the relative external context and therefore could be
ignored for the purposes of those experiments. It all depends on what the
observer is trying to do.

In the real world there is more than one observer, so a theory of
observation should take account of that fact. That is readily done in our
theory. For example, the ground state for two or more distinct observers for
which some commonality of time had been established would be represented by
elements in $\mho $ of the form%
\begin{equation}
|\Psi ^{1},\Psi ^{2})\equiv \mathbb{C}^{1}_{1}\mathbb{C}^{1}_{2}\ldots 
\mathbb{C}^{1}_{r_{1}}\mathbb{C}^{2}_{1}\mathbb{C}^{2}_{2}\ldots \mathbb{C}%
^{2}_{r_{2}}|\Omega ),  \label{yyy}
\end{equation}%
and so on, where superscripts refer to the
different observers. If subsequent dynamics was such that the ESDs of
observer $1$ never sent signals to those of observer $2$ and vice versa,
then to all intents and purposes we could discuss each observer as if they
were alone. If on the other hand some signals did pass between them, then
that would be equivalent to having only one observer.

If no commonality of time or other context has been established between the
observers, then there can be no physical meaning to (\ref{yyy}). This is an
important point in cosmology, where there are frequent discussions about
multiple universe `bubbles' beyond the limits of observation. The mere fact
that astronomers have received light from extremely distant galaxies
establishes a context between the signal preparation ESDs associated with
those galaxies and the ESDs associated with the astronomers now and
validates the use of General Relativity for those regions of spacetime. If
no such signals have been received, then there is no such context.
Therefore, relative to astronomers today, the universe beyond the horizon of
observation can be meaningfully represented only by the information void,
not the spatial vacuum. Something may be there, but we should not discuss it
as if we had access to any form of information about it, such as its
spacetime structure.

Much the same concern must be raised about the loss of information question
in black hole physics. The answer to that question can come only from a
careful understanding of the contextual relation between observers outside
the critical radius and those that were assumed to be inside it.

\section{Experiments}

Long before any experiment can begin, the observer starts off with a
laboratory $\Lambda $ in its void state $|\Omega )$. Then at some time $%
t_{-1}$ before any runs can be taken, specific apparatus consisting of a
finite number $r$ of ESDs has to be constructed in $\Lambda $. We will
assume without loss of generality that these are all functioning normally
and in their ground state, so the labstate $|\Psi , t_{-1})$ at that point is
given by the right-hand side of (\ref{lab}). All of this is necessary before
state preparation.

According to what we said earlier, external context involving ESDs in their
empty state can be ignored. Therefore, we need only discuss those ESDs which
subsequently are in states $|0)$, $|1)$ or $|\mathsf{B})$. A further
simplification is that in real experiments, observers generally filter out
observations from faulty ESDs (assuming these have been identified) by
post-selecting only those labstates which contain the normal bit states $|0)$
or $|1)$. We shall confine our attention to such normal
labstates until we deal with applications to quantum mechanics.

Given this condition, we can restrict our discussion at any given time $%
t_{n} $ to the \emph{physical register $\mathcal{R}_{n}$}, a subset of the
universal register $\mathcal{\mathsf{\mho }}$ consisting of $2^{r_{n}}$
normal states, each of the form 
\begin{equation}
|i_{1}i_{2}\ldots i_{r_{n}})\equiv |i_{1})|i_{2})\ldots |i_{r_{n}}),
\label{labstate}
\end{equation}%
where $i_{j}=0$ or $1$ for $j=1,2,\ldots ,r_{n}$, such that $|i_{j})$ is in $%
\mathcal{P(}\mathsf{B}_{n}^{j})$. The physical register $\mathcal{R}_{n}$
represents all those ESDs in the laboratory $\Lambda $ at time $n$ which
exist and are not faulty.

Given the set of bits $\mathsf{B}_{n}^{j}\equiv \{|0_{j}),|1_{j})\}$
associated with $\mathcal{R}_{n}$, the label $j$ gives an ordering, so $%
\mathcal{R}_{n}$ can be regarded as the Cartesian product $\mathsf{B}%
_{n}^{1}\times \mathsf{B}_{n}^{2}\times \ldots \times \mathsf{B}%
_{n}^{i_{r_{n}}}$. When we come to discuss quantization towards the end of
this paper, every labstate (\ref{labstate}) in $\mathcal{R}_{n}$ will be
identified with a qubit tensor product state $|i_{1}) \otimes
|i_{2}) \otimes \ldots \otimes |i_{r_{n}}) $, an element of the
preferred basis for the associated qubit register.

The notation (\ref{labstate}) will be referred to as the \emph{occupancy}
notation, as the integers $i_{j}$ can be interpreted as the answer to the
question whether the $j^{th}$ ESD ${\Delta }_{j}$ contains a signal or is in
its ground state. An occupancy value $0$ means $\Delta _{j}$ is in its
ground state whilst the occupancy value $1$ means that $\Delta _{j}$ is in
its signal state. These states satisfy the orthonormality conditions $%
(i_{1}i_{2}\ldots i_{r_{n}}|j_{1}j_{2}\ldots j_{r_{n}})=\delta
_{i_{1}j_{1}}\delta _{i_{2}j_{2}}\ldots \delta _{i_{r_{n}}j_{r_{n}}}$.

We define the \emph{signality} of a given state in $\mathcal{R}_{n}$ to be
the number of ones in the occupancy representation of that state. For
example, the state $|00101101)$ is a signality-four state in a rank-$8$
physical register. Signality allows us to partition the $2^{r_{n}}$ states
in $\mathcal{R}_{n}$ into a number of \emph{signal classes }$\mathcal{S}^{0}$%
, $\mathcal{S}^{1},\ldots ,\mathcal{S}^{r_{n}}$. These are equivalence
classes of states in $\mathcal{R}_{n}$ defined by the same signality.

Signality has physical significance. The zero-signal class $\mathcal{S}^{0}$
consists of one state only, the ground state $|000\ldots 0)$ of the physical
register. States in the \emph{one-signal} class $\mathcal{S}^{1}$ correspond
to what would normally be called a one-particle state, states in $\mathcal{S}%
^{2}$ correspond to two-particle states, and so on.

There is a total of $r_{n}+1$ distinct signal classes. The $d^{th}-$signal
class $\mathcal{S}^{d}$ contains $C_{d}^{r_{n}}\equiv r_{n}!/d!\left(
r_{n}-d\right) !$ distinct states. The $r_{n}$-signal class $\mathcal{S}%
^{r_{n}}$ consists of only one state, the \emph{fully saturated\ state} $%
|1_{1})|1_{2})\ldots |1_{r_{n}})$.

Given a rank-$r$ physical register $\mathcal{R}_{r}\ \equiv \mathsf{B}^{1}%
\mathsf{B}^{2}\ldots \mathsf{B}^{r}$ we define the $r$ signal creation
operators%
\begin{equation}
\overline{\mathbb{A}}_{i}\equiv \left\{ \prod_{j\neq i}^{\infty
}I_{j}\right\} \times \bar{A}_{i},\ \ \ \ \ 1\leqslant i\leqslant r,
\end{equation}%
and the $r$ signal annihilation operators%
\begin{equation}
\mathbb{A}_{i}\equiv \left\{ \prod_{j\neq i}^{\infty }I_{j}\right\} \times
A_{i},\ \ \ \ \ 1\leqslant i\leqslant r\text{.}
\end{equation}%
Then an application of the operator $\mathbb{A}_{i}$ on the contextual rank-$r$ ground
state $\mathbb{C}_{1}\mathbb{C}_{2}\ldots \mathbb{C}_{r} | \Omega )$ gives the
rank-$(r-1)$ ground state%
\begin{equation}
\mathbb{A}_{i}\mathbb{C}_{1}\mathbb{C}_{2}\ldots \mathbb{C}_{r} | \Omega ) =%
\mathbb{C}_{1}\mathbb{C}_{2}\ldots \mathbb{C}_{i-1}\mathbb{C}_{i+1}\ldots 
\mathbb{C}_{r} | \Omega ).
\end{equation}%
This is a non-zero labstate in $\mathsf{\mho }$, but is not an element of
the original physical register $\mathcal{R}_{r}\ \equiv \mathsf{B}^{1}%
\mathsf{B}^{2}\ldots \mathsf{B}^{r}$. What has happened is analogous to the
convention qubit register result $a_{i}\{|0_{1}\rangle \otimes |0_{2}\rangle
\otimes \ldots \otimes |0_{r}\rangle \} = 0$. In our case, we do not get
zero, but the equivalent of it: the action of $\mathbb{A}_{i}$ on the ground
state $\mathbb{C}_{1}\mathbb{C}_{2}\ldots \mathbb{C}_{r}| \Omega )$ of $%
\mathcal{R}_{r}$ maps it into the ground state of a different physical
register, one of rank $r-1$, i.e., into a state orthogonal to every state in 
$\mathcal{R}_{r}$.

The signal creation operators $\overline{\mathbb{A}}_{i}$ can be used to
create the various signal classes discussed above, as follows. We start from
the signality-zero class $\mathcal{S}^{0},$ which consists of no application
of any $\overline{\mathbb{A}}_{i}$ to the contextual ground state $|0)\equiv \mathbb{C}%
_{1}\mathbb{C}_{2}\ldots \mathbb{C}_{r} | \Omega )$.

The signality-one class $\mathcal{S}^{1}$ consists of states of the form $%
|2^{i-1})\equiv \overline{\mathbb{A}}_{i}|0)$ for $i=1,2,\ldots ,r,$ the
signality-two class consists of all states of the form $|2^{i-1}+2^{j-1})%
\equiv \overline{\mathbb{A}}_{i}\overline{\mathbb{A}}_{j}|0)$ for $%
1\leqslant i<j\leqslant r$, and so on. Finally, the signality-$r$ signal
class consists of the single state $|2^{r}-1)\equiv \overline{\mathbb{A}}_{1}%
\overline{\mathbb{A}}_{2}\ldots \overline{\mathbb{A}}_{r}|0)$.

In the following discussion, we shall use the notation $|k)$, $k=0,1,\ldots
,2^{r}-1$ for the $2^{r}$ states in $\mathcal{R}^{r}$ and refer to it as the 
\emph{computational basis}.

\section{Classical particle signal mechanics}

In this section, we shall restrict our attention to classical mechanics, to
illustrate how our approach to observation can apply in that context.
Quantum mechanics is discussed after that.

We consider now a physical register $\mathcal{R}^{r}$ of sufficiently large
fixed rank $r$ such that it can serve as a model for a region of classical
physical space over which particles can move. In this approach, particle
motion is discussed in terms of the tracking of signals from a vast
collection of ESDs over time. A particularly useful feature of this approach
is that signality need not be conserved, which means that classical particle
creation and annihilation is readily incorporated into the formalism.

There are several distinct forms of temporal evolution which could be
discussed in such a scenario; the laws of mechanics for a given SUO could
change with time or not, the SUO could be autonomous or interact with
external forces, and signality could be conserved or not. We shall restrict
our attention to autonomous SUOs with time-independent laws of dynamics, as
these are generally of most interest. In principle, there should be no
problem in dealing with other forms of dynamics, including those where the
rank of the physical register changes with time. We could also deal with
classical stochastic mechanics, which would incorporate Bayesian principles
in a natural way.

In the following, all states are elements in the universal register $\mho $
which also belong to $\mathcal{R}^{r}$, i.e., they represent the labstates
of a fixed collection of normal ESDs, each of which can be found only in
either its ground state or signal state.

We shall use the computational basis $\{|k):k=0,1,\ldots ,2^{r}-1\}$ to
represent the $2^{r}$ states in $\mathcal{R}^{r}$. Consider the temporal
evolution of a system from state $|k)$ at time $t$ to state $|Uk)$ after one
elementary time-step, where $k$ and $Uk$ are integers in the interval $%
[0,2^{r}-1]$. Denoting this transition as the action of some temporal
evolution operator $\mathbb{U}$ acting on the initial state $|k)$, we write $%
\mathbb{U}|k)=|Uk)$, $0\leqslant k,Uk<2^{r}$.

For a given $k$, there are in principle $2^{r}$ possible states $|Uk)$ into
which it could be mapped, and because there are $2^{r}$ values of $k$, we
conclude that for a rank-$r$ classical register, there are $(2^{r})^{2^{r}}$
distinct possible evolution operators in this form of mechanics. Even for
very low rank physical registers, the number of possible operators is
impressive. For example, a rank-$2$ register can have $256$ different forms
of autonomous, time-independent dynamics whilst a rank-$3$ register has $%
8^{8} = 16,777,216$ different forms.

Most of the possible evolution operators over a physical register will not
be useful. Many of them will correspond to irreversible and/or unphysical
dynamical evolution and only a small subset will be of interest. We need to
find some principles to guide us in our choice of evolution operator.

Recall that in standard classical mechanics, Hamilton's equations of motion
lead to Liouville's theorem. This tells us that as we track a small volume\
element along a classical trajectory, this volume remains constant in
magnitude though not necessarily constant in shape or orientation. This
leads to the idea that a system of many non-interacting particles moving
along classical trajectories in phase-space behaves like an incompressible
fluid, such a phenomenon being referred to as a Hamiltonian flow.

An important characteristic of Hamiltonian flows is that flow lines never
cross. We shall encode this idea into our approach to signal mechanics.
There are two versions of this mechanics, one of which does not necessarily
conserve signality whilst the other does. We consider the first one now.

\subsubsection{Permutation flows}

The physical register $\mathcal{R}^{r}$ contains $2^{r}$ labstates denoted
by $|k)$, $k=0,1,2,\ldots ,2^{r}-1$. Consider a permutation $P$ of the
integers $k$, such that under $P$, $k\rightarrow Pk\in [ 0,2^{r}-1]$. Define
the evolution of the labstate $|k)$ over one time step by $|k)\rightarrow 
\mathbb{U}|k)=|Pk)$. Such a process is reversible and will be referred to as
a \emph{permutation flow}.

There is a total of $n!$ distinct permutations of n objects, so there are ($%
2^{r})!$ possible distinct permutation flow processes. For large $r$, the
number of permutation flows is a rapidly decreasing fraction of the number ($%
2^{r})^{2^{r}}$ of all possible forms of register processes.

\subsubsection{Signal conserving flows}

Most permutation flows will not conserve signality. We can readily identify
the subset of the permutation flows which do conserve signality by using the
occupancy notation. Consider a physical register state $|\Psi _{n})$ at time 
$t_{n}$ given by $|\Psi _{n})=|i_{1}i_{2}\ldots i_{r})$ in the occupancy
notation, where $i_{j}=0$ or else $1$ for $1\leqslant j\leqslant r$.

Now let $P^{\ast }$ be some permutation of the numbers $1,2,\ldots ,r$ and
write $P^{\ast }j$ to represent the number that $j$ changes to under this
permutation. Now suppose that $|\Psi _{n})$ evolves into the labstate $|\Psi
_{n+1})$ at time $t_{n+1}$ given by 
\begin{equation}
|\Psi _{n})\rightarrow |\Psi _{n+1})\equiv \mathbb{U}|\psi )=|i_{P^{\ast
}1}i_{P^{\ast }2}\ldots i_{P^{\ast }r}).
\end{equation}%
To determine the new occupancy of the $j^{th}$ bit, we just look at the
occupancy of the ($P^{\ast }j)^{th}$ bit. This may be summarized as the
dynamical rule $i_{j}\rightarrow i_{j}^{\prime }\equiv i_{P^{\ast }j}$. We
shall call this form of signal mechanics \emph{signal permutation dynamics.}

In this form of dynamics, signality is automatically conserved. Another way
of seeing this is to use the signal creation operators and note that if $%
|\Psi _{n})$ has signality $d$, then we can write $|\Psi _{n})=\overline{%
\mathbb{A}}_{j_{1}}\overline{\mathbb{A}}_{j_{2}}\ldots \overline{\mathbb{A}}%
_{j_{d}}|0)$, where $1\leqslant j_{1}<j_{2}<\ldots <j_{d}\leqslant r$. Then
under the above permutation $P^{\ast }$ of the integers $1,2,\ldots ,r$ the
new state at time $t_{n+1}$ takes the form%
\begin{equation}
|\Psi _{n+1})\equiv \mathbb{U}|\Psi _{n})=\overline{\mathbb{A}}_{P^{\ast
}j_{1}}\overline{\mathbb{A}}_{P^{\ast }j_{2}}\ldots \overline{\mathbb{A}}%
_{P^{\ast }j_{d}}|0).
\end{equation}%
Then clearly signality is conserved.

The total number of distinct permutations of $r$ objects is $r!$, so there
are that many distinct forms of signal permutation dynamics for a rank-$r$
classical register. Since there are $\left( 2^{r}\right) !$ distinct forms
of permutation dynamics, the set of signal permutation dynamics forms a
rapidly decreasing fraction of the set of all possible permutation dynamics.

Permutation flows have a number of features which have analogues in standard
classical mechanics. First, permutation flows are reversible. Given a
permutation $P$, then its inverse $P^{-1}$ always exists, because
permutations form a group.

Another feature of permutation dynamics is the existence of \emph{orbits} or 
\emph{cycles. }A\emph{\ }permutation of $2^{r}$ objects will in general
contain cycles, which are subsets of the objects such that only elements
within a given cycle replace each other under the permutation. This is
relevant here because we have chosen to discuss time-independent autonomous
systems, the evolution of which is given by repeated applications of the
same permutation. Therefore, the structure of the cycles does not change and
so each cycle consisting of $p$ elements has a dynamical period $p$. For
example, the identity permutation gives a trivial form of mechanics where
nothing changes. It has $2^{r}$ cycles each of period $1$. At the other end
of the spectrum, the permutation denoted by $(0\rightarrow 1\rightarrow
2\rightarrow \ldots \rightarrow 2^{2}-1\rightarrow 0)$ has no cycles except
itself and has period $2^{r}$. Therefore, any physical register evolving
under time independent, autonomous permutation mechanics must return to its
initial labstate no later than after $2^{r}$ time steps. This is the
analogue of the Poincar\'{e} recurrence theorem \cite{POINCARE-1890}.

\section{Evolution and measurement}

Any experiment consists of several distinct phases. We have discussed the
creation of the apparatus and the evolution of the labstates. Now we turn to
the process of measurement itself, which denotes the extraction of classical
information from an SUO. Typically this information will be in the form of
real numbers, and these can always be expressed in binary form, justifying
our approach.

Context plays a vital role here. When for example an observer reports that a
particle has been observed at position $x = 1.5$, what they mean is that
positive signals have been detected at some normal ESD or ESDs associated
with the number $x = 1.5$. This assignment is based on the context of the
experiment: the observer will know on the basis of prior theoretical
knowledge what those ESDs mean in terms of the physics of the SUO concerned,
and therefore, what $``$values$"$ of some measurable quantity they represent.

So far we have discussed the evolution on labstates. For each run or
repetition of the experiment, this is modeled by the action of an evolution
operator $\mathbb{U}_{N}$ mapping initial labstates at time $t_{0}$ into
final labstates at time $t_{N}$. We need now to discuss how numbers are
extracted at the end of an experiment consisting of a number of runs.

With reference to the position measurement discussed immediately above, we
model the measurement process in terms of \emph{weighted relevant questions}%
. What this means is this. Suppose the final physical register $\mathcal{R}%
_{N}$ has rank $r_{N}$. Assuming the experimentalist has established that
each ESD is normal, then there will be a total of $d_{N}\equiv 2^{r_{N}}$
possible normal labstates in this register. Therefore, the observer could
ask a total of d$_{N}$ normal questions. These questions are represented by
the dual labstates $\left\{ (k|:k=0,1,\ldots ,d_{N}-1\right\} $. Given a
final labstate $|\Psi _{N})$, the answer \emph{yes} or \emph{no} to each
question $\ (k|\ \equiv $ $``$\emph{is it true that }$|\Psi _{N})$ is $|k)$%
\emph{?}$"$ is represented by the number one or zero respectively, and given
by the bracket $(k|\Psi _{N})$.

Now the observer will generally have some theory as to what each answer $|k)$
means physically. In many experiments, this will be some real number $X_{k}$%
. Therefore, the actual number $\langle X\rangle _{\Psi _{N}}$ obtained at
time $t_{N}$ at end of a single run of the experiment can be written in the
form%
\begin{equation}
\langle X\rangle _{\Psi _{N}}=(\Psi _{N}|\mathbb{X}_{N}|\Psi _{N}),
\label{ask}
\end{equation}%
where $\mathbb{X}_{N}\equiv \sum_{k=0}^{d_{N}-1}|k)X_{k}(k|$ is an \emph{%
observable}, a sum of dyadics representing a weighted relevant question.

Two comments are relevant. First, despite appearances, this is still a
classical theory at this point. The final labstate $|\Psi _{N})$ is a single
element in the final physical register, $\mathcal{R}_{N}$, not a
superposition of elements. Second, there is nothing in classical mechanics
which rules out weighted sums of dyadics. For any element in $\mathcal{R}%
_{N} $, all the possible answers $(k|\Psi _{N})$ are zero except for one of
them, so (\ref{ask}) returns a physically sensible value for $\langle
X\rangle _{\Psi _{N}}$.

A further refinement, anticipating the possibility of random variations in
the initial state and the extension of these ideas to quantum mechanics is
to write%
\begin{equation}
\langle X\rangle _{\Psi _{N}}=Tr\{\mathbb{X}_{N}\rho _{N}\},  \label{av}
\end{equation}%
where $Tr$ represents the familiar trace process and $\rho _{N}$ is the
dyadic $|\Psi _{N})(\Psi _{N}|$ analogous to a pure state density matrix in
quantum mechanics.

We note that $|\Psi _{N})=\mathbb{U}_{N,0}|\Psi _{0})$ and $(\Psi _{N}|\
=(\Psi _{N}|\overline{\mathbb{U}}_{N,0}$, where the evolution operator $%
\mathbb{U}_{N,0}$ maps elements of $\mathcal{R}_{0}$ into elements of $%
\mathcal{R}_{N}$ and similarly for the dual evolution operator $\overline{%
\mathbb{U}}_{N,0}$. In general, it will be true that 
\begin{equation}
\overline{\mathbb{U}}_{N,0}\mathbb{U}_{N,0}=\mathbb{I}_{\mathcal{R}_{0}}%
\text{,}  \label{semi}
\end{equation}%
the identity operator for $\mathcal{R}_{0}$. However, because there is no
requirement formally in this approach for the rank $r_{N}$ of the final
physical register $\mathcal{R}_{N}$ to equal the rank $r_{0}$ of the initial
physical register $\mathcal{R}_{0}$, it is possible that $\mathbb{U}_{N,0}%
\overline{\mathbb{U}}_{N,0}$ does not equal $\mathbb{I}_{\mathcal{R}_{N}}$.
This corresponds to irreversible dynamics. In the analogous quantum
formalism we have developed \cite{J2008R}, such evolution operators
are referred to as \emph{semi-unitary}.

Using (\ref{semi}) in (\ref{av}), we may write $\langle X\rangle _{\Psi
_{N}}=Tr\{\mathbb{X}_{N}\mathbb{U}_{N,0}\rho _{0}\overline{\mathbb{U}}%
_{N,0}\}$, where $\rho _{0}$ is the initial dyadic $|\Psi _{0})(\Psi _{0}|$.

\subsection{Random initial states}

Real experiments normally consist of a large number of repetitions or runs
of a basic process. However, it cannot always be guaranteed that the initial
labstate is always the same. In principle, we could start with any one of $%
d_{0}\equiv 2^{r_{0}}$ initial labstates. In such a case, a statistical
approach can be taken.

Consider a very large number $R$ of runs, such that there is a total of $%
R_{k}$ runs starting with initial labstate $|k)$, for $k=0,1,\ldots ,d_{0}-1$%
. Clearly, $\sum_{k=0}^{d_{0}-1}R_{k}=R$. Then in the limit of $R$ tending
to infinity, we would assign a probability $\omega _{k}\equiv
\lim_{R\rightarrow \infty }R_{k}/R$ for the initial labstate to be in state $%
|k)$.

In such a scenario we define the initial density matrix $\rho _{0}\equiv \sum_{k=0}^{d_{0}-1}\omega
_{k}|k,0)(k,0|$, where $|k,0)$ is any one of the $d_{0}$ elements of the
initial physical register $\mathcal{R}_{0}$ and the $\omega _{k}$ are
probabilities summing to unity. The formalism outlined above then gives the expectation values of operators.

\section{Quantization}

The formalism we have developed is readily extended to the quantum scenario,
for which different principles hold concerning the interpretation and usage
of the physical register. In this scenario, a rank $r_{n}$ physical register 
$\mathcal{R}_{n}$ at time $n$ is identified with a preferred orthonormal
basis $\{|k):k=0,1,\ldots ,d_{n}-1\}$ for a quantum register $\mathcal{Q}%
_{n}\equiv Q^{1}\otimes Q^{2}\otimes \ldots \otimes Q^{r_{n}}$, where now
the $Q^{i}$ are qubits. This register is a Hilbert space, the tensor product
of $r_{n}$ qubits, each of which is identified with one ESD. Elements of
this register are the labstates of interest and these can be multiplied by
complex numbers and added together, unlike the classical scenario. Most of
the formulae developed in the previous section can be taken wholesale into
the quantum scenario. For example, (\ref{ask}) now corresponds to the
expectation value of Hermitian operator $\hat{X}\equiv
\sum_{k=0}^{d_{N}-1}|k)X_{k}(k|$ relative to the normalized pure quantum
state $|\Psi )=\sum_{k=0}^{d_{N}-1}\Psi _{k}|k)$, where $%
\sum_{k=0}^{d_{N}-1}|\Psi _{k}|^{2}=1.$

In applications to quantum optics, this approach has been extended to
include SUO attributes such as internal spin \cite{J2008E,J2008F} in
a generalization of the Ludwig-Kraus POVM formalism, which extends the work
of von Neumann on quantum measurement \cite{VON-NEUMANN-1955}. Properties such as
spin and electric charge, conventionally interpreted as objective properties
of SUO states, are encoded as contextual properties of ESDs. This is a
realization of one of the aims of Feynman's thesis, in which he wrote \cite{BROWN-2005}: 
`\emph{and all of the apparent quantum properties of light and the
existence of photons may be nothing more than the result of matter
interacting with matter directly, and according to quantum mechanical laws.}'

In the next two sections we show how our formalism describes experiments
where the apparatus changes in one way or another during the experiment. In
particular, the faulty state plays an important role in these
experiments.

\section{The Elitzur-Vaidman bomb-tester experiment}

In this experiment, a stockpile of active (A) and dud (D) bombs is analysed,
one by one, in order to find as many unexploded active bombs as possible.
The approach follows that discussed in \cite{ELITZUR+VAIDMAN-1993}. In the
schematic Mach-Zehnder circuit shown in Figure 1, each numbered circle
represents an ESD, or place where the observer could extract information in
principle. The rectangles represent beam-splitters and the solid bars represent mirrors. Our convention is that a
reflected beam undergoes a phase change of $\pi $ whilst a transmission
results in a phase change of $\pi /2$. For each run of the experiment, a
bomb is taken from the untested stockpile of bombs and is placed in contact
with ESD\ 2, such that if the bomb is a dud (D) then ESD 2 acts normally and
transmits onto ESD 5. If on the other hand the bomb is active (A), then whenever the
signal state is triggered at ESD2 the bomb explodes. In this case, ESD
2 becomes \emph{faulty} and does not transmit  onto ESD 5.

\begin{figure}[htbp]
\centering
\includegraphics[width=3.5in]{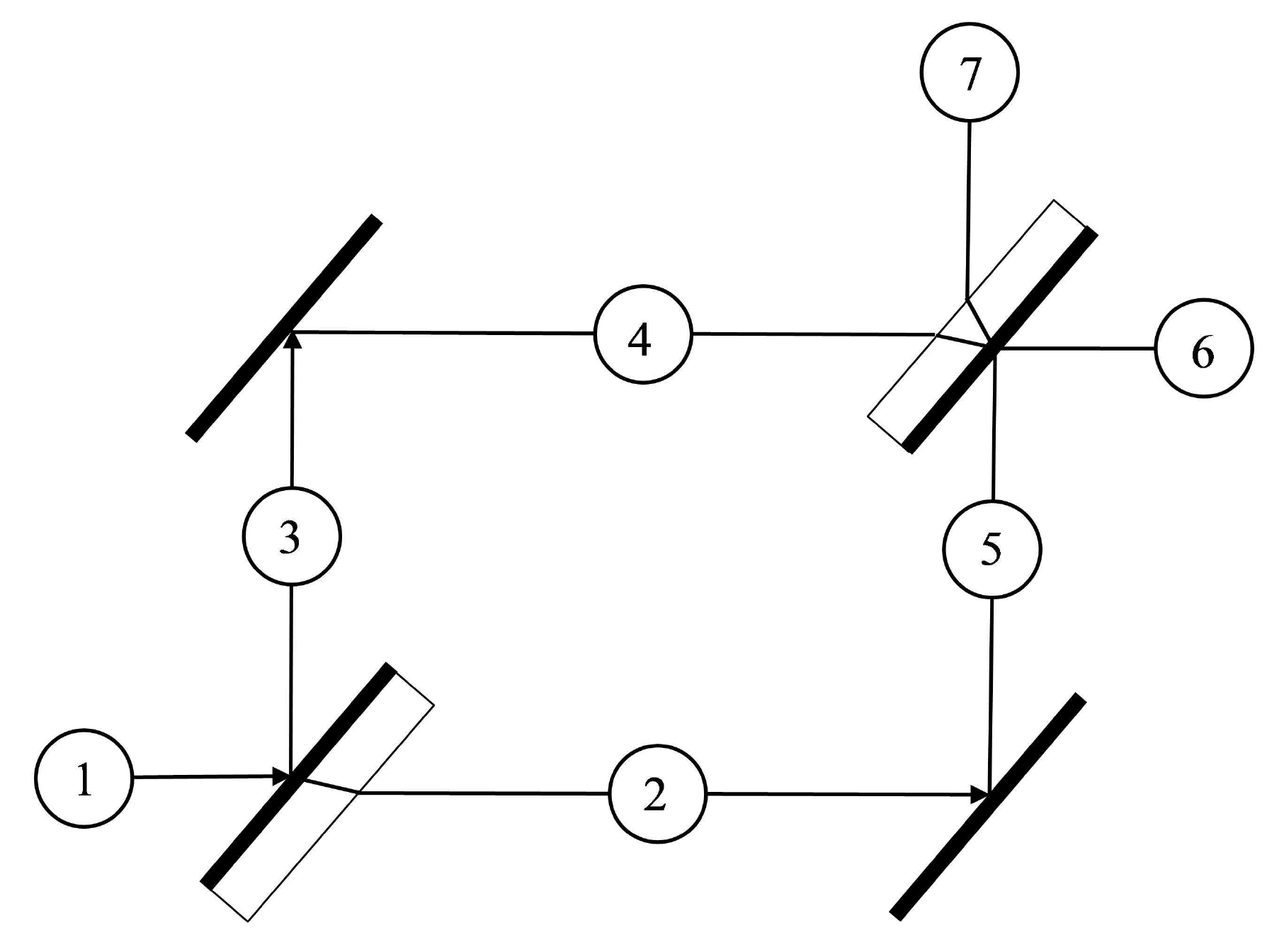}
\caption{The Mach-Zehnder network.}
\end{figure}

This experiment involves a randomly changing apparatus network, because the bomb 
being tested in a given run is immediately replaced by a new bomb from the 
unused stockpile. There are therefore two distinct networks to consider. We 
have to deal with each of these separately, for which a pure labstate description 
can be used.  Then we use a density matrix approach to consider the combined experiment.

For each of the pure labstate calculations, the
labstate, or current signal state of the network, is discussed in the Schr\"{o}dinger picture. At any stage 
$n$, the current labstate will be written in the form $%
|\Psi ,n)$. We define the contextual vacuum $|0)\equiv \mathbb{C}_{1}\mathbb{C}_{2}\mathbb{C}_{3}\mathbb{C}_{4}\mathbb{C}%
_{5}\mathbb{C}_{6}\mathbb{C}_{7}|\Omega )$, where $\mathbb{C}_{i}$ is the construction operator for the $i^{th}$ ESD.

\subsection{Networks with a dud bomb}

Starting at time $t=0$, the initial labstate $|D,0) \equiv \overline{\mathbb{A}}_{1}|0)$ 
changes by the following sequence of stages, or opportunities for information extraction from the apparatus: 
\begin{eqnarray}
|D,0)\rightarrow |D,1) &=& \frac{1}{\sqrt{2}}\{i\overline{\mathbb{A}}_{2}-%
\overline{\mathbb{A}}_{3}\}|0) \nonumber \\
|D,1) \rightarrow |D,2) &=& \frac{1}{\sqrt{2}}\{-i\overline{\mathbb{A}}_{5}+%
\overline{\mathbb{A}}_{4}\}|0).
\end{eqnarray}%
Using 
\begin{equation}
\overline{\mathbb{A}}_{4}|0)\rightarrow \frac{1}{\sqrt{2}}\{i\overline{%
\mathbb{A}}_{6}-\overline{\mathbb{A}}_{7}\}|0),\ \ \ \overline{\mathbb{A}}%
_{5}|0)\rightarrow \frac{1}{\sqrt{2}}\{-\overline{\mathbb{A}}_{6}+i\overline{%
\mathbb{A}}_{7}\}|0)
\end{equation}%
we find  
$|D,2)\rightarrow |D,3)=\overline{\mathbb{A}}_{6}|0)$. Hence no dud bomb ever coincides with a signal in ESD 7.

\subsection{Networks with an active bomb}

In this case, the initial labstate is given by
$|A,0) \equiv \overline{\mathbb{A}}_{1}\mathbb{D}_{2}|0)$,
where $\mathbb{D}_{i}$ is the decommissioning operator for ESD i.
In this case we have
\begin{eqnarray}
|A,0)\rightarrow |A,1) &=& \frac{1}{\sqrt{2}}\{i\mathbb{D}_{2}-\overline{\mathbb{%
A}}_{3}\}\mathbb{D}_{2}|0), \nonumber \\
|A,1)\rightarrow |A,2) &=& \frac{1}{\sqrt{2}}\{i\mathbb{D}_{2}+%
\overline{\mathbb{A}}_{4}\}\mathbb{D}_{2}|0) , \\
|A,2)\rightarrow |A,3) &= & \frac{1}{2}\{\sqrt{2}i\mathbb{D}_{2} +
i\overline{\mathbb{A}}_{6}-\overline{\mathbb{A}}_{7}\} \mathbb{D}_{2}|0).
\end{eqnarray} 
Using $\mathbb{D}_{2}\mathbb{D}_{2} = \mathbb{D}_{2}$, the state $\mathbb{D}_{2}|0)$ 
represents an explosion at ESD 2 coincident with no signal at either ESD 6 or 7. Hence this 
calculation gives the outcome probabilities
$P(\text{Explode}|A)=\frac{1}{2}$, $P(6|A)=P(7|A)=\frac{1}{4}$.

\subsection{Random testing}

Unfortunately, the observer does not know before each run whether a particular bomb 
is active or a dud. Consider a sequence of runs such that there is a (classical) probability $\omega
_{A} $ of encountering an active bomb and a probability $\omega
_{D}=1-\omega _{A} $ of a dud. In this case we take a density matrix
approach. At the $n^{th}$  stage we define the density matrix
\begin{equation}
\rho (n)=\omega _{A}|A,n)(A,n|+\omega_{D} |D,n)(D,n|.
\end{equation}%
The overall probability of triggering ESD 6 at the third stage is given by%
\begin{equation}
P(6|\rho ,3)=Tr\{\rho (3)\mathbb{P}^{1}_{6}\}=\frac{\omega _{A}}{4}+\omega
_{D},
\end{equation}
where $\mathbb{P}^{1}_{i}$ is the signal projection operator for the $i^{th}$ ESD, and likewise,
\begin{equation}
P(7|\rho ,3)=Tr\{\rho (3)\mathbb{P}^{1}_{7}\}=\frac{\omega _{A}}{4}.
\end{equation}
The interpretation is that every signal at ESD 7 tells the observer that the current bomb on test is
active. Moreover, that bomb has not exploded and can be used. This method allows 
one quarter of the active bombs in the stockpile to be identified with a single 
sweep of the stockpile. Further iteration of this experiment on unexploded bombs 
coinciding with a signal from ESD 6 pushes this result up to one
third of the total, the rest of the active bombs having exploded during testing.

\section{The Hardy paradox experiment}

The Elitzur-Vaidman bomb-tester experiment may be interpreted as a simplified form of 
double-slit experiment, where the screen has only two sites and one of the slits can be 
blocked off or not, depending on whether a bomb is active or dud. This blocking off 
occurs in a classical way, because the uncertainty as to whether the bomb is active or 
dud is not intrinsic to the nature of the bomb but reflects the observer's ignorance of the nature of the bomb.

A spectacular variant of the bomb-tester experiment is known as the Hardy paradox 
experiment \cite{HARDY-1992}. In this variant, the blocking off of a slit occurs in an intrinsically 
quantum stochastic way, contrasted to the classically stochastic way of the bomb-tester 
experiment. The experiment consists of an electron-positron pair passing through 
two coupled Mach-Zehnder-type networks shown in Figure 2. The curvature of the 
tracks occurs because of suitable magnetic fields perpendicular to the plane of the network shown.

\begin{figure}[htbp]
\centering
\includegraphics[width=3.5in]{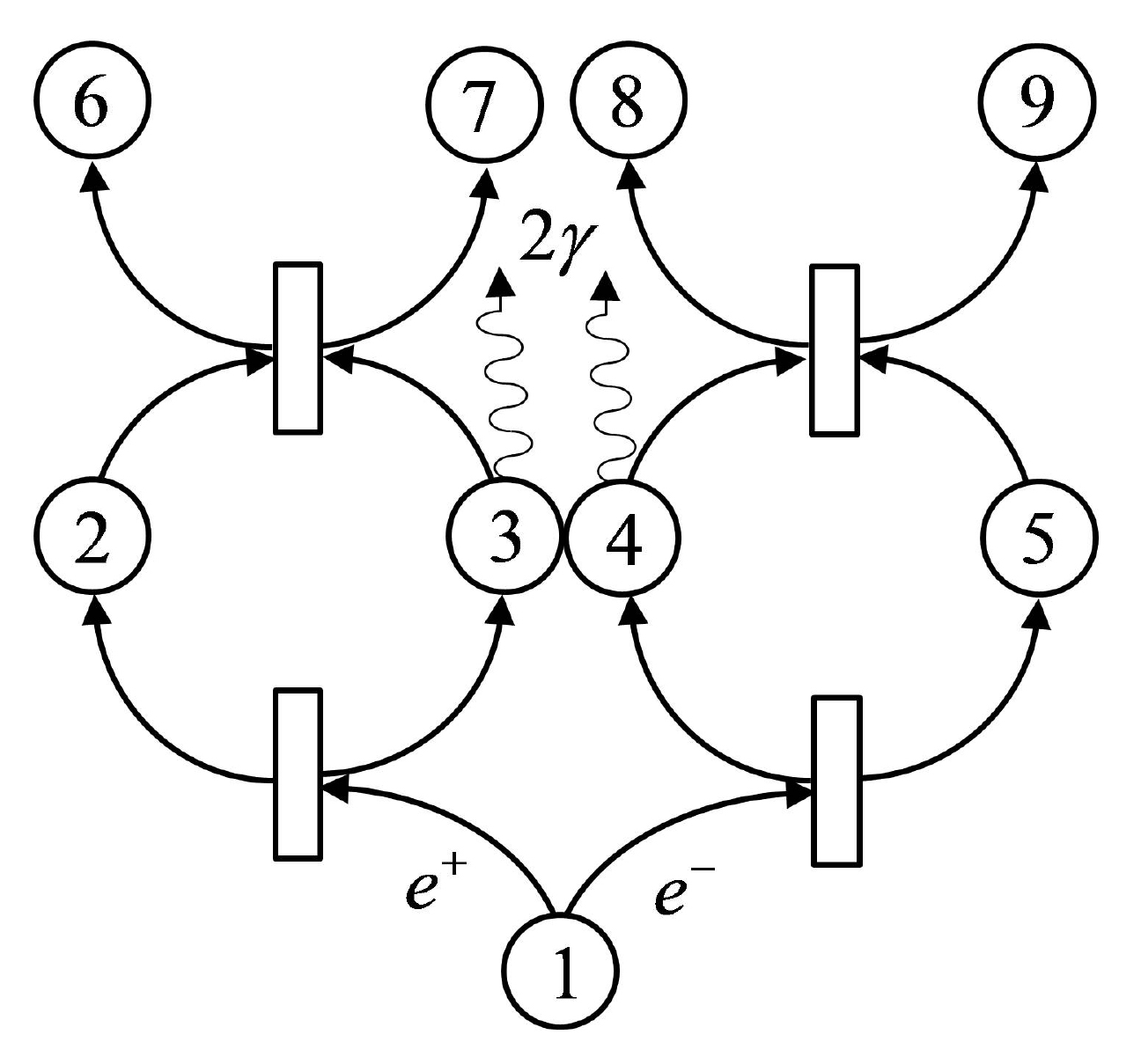}
\caption{The Hardy paradox network.}
\end{figure}

In conventional terminology, the presence of the positron at ESD 3 would effectively 
block one of the slits, i.e., the slit corresponding to ESD 4, through which the electron 
wavefunction would otherwise pass. Conversely, the absence of the positron at ESD 3 allows 
both slits 4 and 5 to be open as far as the electron is concerned. By symmetry, the same 
remarks apply to the interchange of the electron and positron.

The Hardy paradox experiment is intrinsically a pure quantum experiment, in that we can 
discuss it via pure labstates alone. An important point is that electron-positron annihilation 
is a well-known quantum process which occurs in nature, whereas the detonation mechanism 
of the Eliztur-Vaidman bomb-tester is left unspecified. 

We start our analysis of the Hardy paradox experiment by defining the contextual vacuum 
$|0)\equiv \mathbb{C}_{1}\mathbb{C}_{2}\mathbb{C}_{3}\mathbb{C}_{4}\mathbb{C}%
_{5}\mathbb{C}_{6}\mathbb{C}_{7}\mathbb{C}_{8}\mathbb{C}_{9}|\Omega )$.
Then the initial state is given by $|\Psi ,0)\equiv \overline{\mathbb{A}}_{1}|0)
$.
The dynamics follows the following sequence:
\begin{equation}
|\Psi ,0)\rightarrow |\Psi ,1)=\frac{1}{2}(i\overline{\mathbb{A}}_{2}-%
\overline{\mathbb{A}}_{3})(i\overline{\mathbb{A}}_{5}-\overline{\mathbb{A}}%
_{4})|0).
\end{equation}

Using the labstate evolutions
\begin{equation}
\begin{array}{rl}
\overline{\mathbb{A}}_{2}\overline{\mathbb{A}}_{4}|0)\rightarrow  & \frac{1}{%
2}(i\overline{\mathbb{A}}_{7}-\overline{\mathbb{A}}_{6})(i\overline{\mathbb{A%
}}_{9}-\overline{\mathbb{A}}_{8})|0) \\ 
\overline{\mathbb{A}}_{2}\overline{\mathbb{A}}_{5}|0)\rightarrow  & \frac{1}{%
2}(i\overline{\mathbb{A}}_{7}-\overline{\mathbb{A}}_{6})(i\overline{\mathbb{A%
}}_{8}-\overline{\mathbb{A}}_{9})|0) \\ 
\overline{\mathbb{A}}_{3}\overline{\mathbb{A}}_{4}|0)\rightarrow  & \mathbb{D%
}_{3}\mathbb{D}_{4}|0) \\ 
\overline{\mathbb{A}}_{3}\overline{\mathbb{A}}_{5}|0)\rightarrow  & \frac{1}{%
2}(i\overline{\mathbb{A}}_{6}-\overline{\mathbb{A}}_{7})(i\overline{\mathbb{A%
}}_{8}-\overline{\mathbb{A}}_{9})|0)%
\end{array}%
\end{equation}%
we find
\begin{eqnarray}
|\Psi ,1)\rightarrow |\Psi ,2)=\frac{1}{4}\{2\mathbb{D}_{3}\mathbb{D}_{4}+i%
\overline{\mathbb{A}}_{6}\overline{\mathbb{A}}_{8}-\overline{\mathbb{A}}_{7}%
\overline{\mathbb{A}}_{8} \\ \nonumber
-3\overline{\mathbb{A}}_{6}\overline{\mathbb{A}}%
_{9}+i\overline{\mathbb{A}}_{7}\overline{\mathbb{A}}_{9}\}|0).
\end{eqnarray}
The state $\mathbb{D}_{3}\mathbb{D}_{4}|0)$ is interpreted as the occurrence 
of electron-positron annihilation, as shown by the wavy lines in Figure 2,
so we read off the following conditional probabilities:
\begin{eqnarray}
P(6,8| \Psi ) &=& P(7,8| \Psi )=P(7,9| \Psi )=\frac{1}{16} \\ \nonumber
P(6,9| \Psi ) &=& \frac{9}{16}, \ \ \  P(\text{Explosion}| \Psi ) = \frac{1}{4}.
\end{eqnarray}%
The paradox is that $P(7,8| \Psi )\neq 0$, which conventional logical suggests 
could occur only if both particles had passed through ESDs 3 and 4 simultaneously 
\emph{without} annihilation, contrary to expectation.

\section{Implications and Comments}

The application of our quantized model of observation to the Elitzure-Vaidman bomb-tester 
and Hardy paradox experiments demonstrates that the concept of faulty or 
decommissioned states has physical significance. 

This paper was motivated in part by a dissatisfaction with conventional
approaches to quantum observation: in our view, too much attention is paid
to states of SUOs and too little to the context of their observation.
Despite the calculational successes of the conventional approach, it seems
to us that there is something deep still missing throughout physics, viz., a
comprehensive dynamical theory of observation. Such a theory should be able to clarify
the relationship between the SUO, observer and apparatus concepts, which may
lead to the resolution of long standing issues in quantum mechanics. We hope
that this paper may be of some value in suggesting new lines of research in
this area.

\end{document}